\documentclass[apj]{emulateapj}
\usepackage{epsfig}
\usepackage{epstopdf}
\usepackage{color} 
\usepackage{natbib}
\usepackage{amssymb}
\usepackage{bm}
\usepackage{rotating}
\usepackage{graphicx}
\usepackage{longtable}
\usepackage[breaklinks,colorlinks, urlcolor=blue,citecolor=blue,linkcolor=blue]{hyperref}
\usepackage{xcolor}
\usepackage{graphicx}
\usepackage{hyperref}
\usepackage{tikz}
\usepackage{soul}
\usepackage{footmisc}
\usepackage{multirow}
\def\rrm{rad m$^{-2}$}

\def\ewr{$EW_{r}$}
 
\def\mgii{Mg~{\sc ii}~} 
\def\mgiia{Mg~{\sc ii}$\lambda$2796} 
\def\mgiib{Mg~{\sc ii}$\lambda$2803}

\def\feiia{Fe~{\sc ii}$\lambda$2600} 
\def\apjl{{ApJ}}%
\def\apjs{{ApJS}}%
\def\aap{{A\&A}}%
\shorttitle{Residual rotation measure \& Mg II absorber}
\shortauthors{Malik, Chand \& Seshadri}

\begin{document}

\title{Role of intervening \mgii absorbers on the rotation measure and fractional polarisation of the background quasars}

\author{Sunil Malik$^{1}$,
 Hum Chand$^2,3$,
  T. R. Seshadri$^1$}

\affil{$^1$ Department of Physics and Astrophysics, University of Delhi, Delhi$-$ 110007, India \\
$^2$ Aryabhatta Research Institute of Observational Sciences (ARIES),
  Manora Peak, Nainital$-$ 263002, India\\
$^3$ Department of Physics and Astronomical Sciences,
Central University of Himachal Pradesh (CUHP), Dharamshala$-$176215, India.   \url{sunilmalik1993@gmail.com}\\ }

\begin{abstract}

We probed the magnetic fields in high-redshift galaxies using  excess extragalactic contribution to residual rotation measure (RRM) for quasar sightlines with intervening Mg {\sc ii} absorbers. Based on a large sample of 1132 quasars,  we have computed RRM distributions broadening using median absolute deviation from mean  ($\sigma^{md}_{rrm}$), and found it to be 17.1$\pm0.7$ rad m$^{-2}$  for 352 sightlines having Mg {\sc ii} intervening absorbers in comparison to its value of $15.1\pm0.6$ rad m$^{-2}$ for 780 sightlines without such absorbers, resulting in an excess broadening ($\sigma_{rrm}^{ex}$) of $8.0\pm1.9$ rad m$^{-2}$  among these two subsamples. This value of $\sigma_{rrm}^{ex}$, has allowed us to constrain the average strength of magnetic field (rest frame) in high redshift galaxies responsible for these Mg {\sc ii} absorbers, to be $\sim 1.3\pm0.3 \mu G$ at a median redshift of 0.92. This estimate of magnetic field is consistent with the reported estimate in earlier studies based on radio-infrared correlation and energy equipartition for galaxies in local universe. A similar analysis on subsample split based on the radio spectral index, $\alpha$, (with $F_{\nu}\propto \nu^{\alpha}$) for flat ($\alpha$ $\geq -0.3$; 315 sources) and steep ($\alpha$ $\leq -0.7$; 476 sources) spectrum sources shows a significant $\sigma_{rrm}^{ex}$ (at 3.5$\sigma$ level) for the former and absent in latter. An anti-correlation found between the $\sigma^{md}_{rrm}$ and percentage polarisation ($p$) with similar Pearson correlation of $-0.62$ and $-0.87$ for subsample with and without Mg {\sc ii}, respectively, suggests main contribution for decrements in the $p$ value to be intrinsic to the local environment of quasars.

\end{abstract}


\section{Introduction}
\label{introduction}
Magnetic field has a significant impact on several astrophysical processes such as transport and confinement of cosmic rays, star formation in the galaxy, cloud collapse, and galactic outflows \citep[e.g.,][]{Mestel1984A&A...136...98M,Rees1987QJRAS..28..197R,1991MNRAS.248...58W,2018Galax...6..142V}.
%\textcolor{red}{To gauge the effect that magnetic field could have on these processes, we need to study the strength and morphology of the magnetic fields at different scales.}
Several probes, such as, synchrotron radiation, Faraday rotation, Zeeman splitting, dust polarisation and dust emission, are available for investigating it  at different length scales. Among them, Faraday rotation (FR) is a powerful probe to study the strength of the line-of-sight component of the magnetic field over cosmic distances
%The contribution to FR comes from the integrated value of the line-of-sight magnetic field along the path of sightline, weighted by the electron number density
~\citep[e.g.,][]{Kronberg1976Natur.263..653K,Kronberg1977A&A....61..771K,Kronberg1982ApJ...263..518K,Welter1984ApJ...279...19W,You2003AcASn..44S.155Y,2005A&A...444..739B,Kronberg2008ApJ...676...70K,Bernet2008Natur.454..302B,Bernet2010ApJ...711..380B,Bernet2012ApJ...761..144B,Hammond2012arXiv1209.1438H,Bhat2013MNRAS.429.2469B}. FR is quantified by the observed rotation measure (RM) which for a radio source at emission redshift, $z_{emi}$, is given by,

\begin{equation}
\label{eq:defRM}
RM(z_{emi}) = 8.1\times 10^{5} \int\limits_{z_{emi}}^{0} \frac{n_e(z)B_\parallel(z)}{(1+z)^{2}}\frac{dl}{dz}dz.
\end{equation}

Here, RM is measured in units of ~\rrm, $n_{e}$ is the number density of electron (in  $cm^{-3}$), $B_\parallel$ is the longitudinal magnetic field component (measured in Gauss) and $dl/dz$ is the column length (in parsecs) per unit redshift interval.

 The total RM can be expressed as a sum of the following four components:
\begin{equation}
RM = RM_{QSO} +RM_{IGM}+RM_{abs} +GRM.
\label{rm}
\end{equation}
Here, $RM_{QSO}$ is the intrinsic contribution due to the source (i.e., the quasar) itself, $RM_{abs}$ denotes the contribution to rotation measure of any intervening galaxy which happens to be in the line of sight, GRM refers to the galactic rotation measure by our galaxy Milky Way, and $RM_{IGM}$ is the contribution by the intergalactic medium (IGM) which is likely to be negligible as compared to that of the other three components.

To study the extra galactic contribution, GRM is a contamination and needs to be removed. On doing this we are left with  the residual rotation measure (RRM) given by,

\begin{equation}
RRM = RM - GRM.
\label{rrm}
\end{equation}

 Further, distribution of RRM in our sample can be divide into two subsamples, one consisting of sightlines with absorption due to one or more intervening galaxies, detected in the form of \mgii absorption systems (i.e., $RM_{QSO} +RM_{IGM}+RM_{abs}$), and  the other with out any such absorption systems (i.e., $RM_{QSO} +RM_{IGM}$). The statistical properties of RRM in these two sub-samples, or more specifically  the difference among them, can be a unique tool to probe the global properties of the magnetic fields at high redshift galaxies ~\citep[e.g.,][]{1995ApJ...445..624O,Kronberg2008ApJ...676...70K,Bernet2008Natur.454..302B,Bernet2010ApJ...711..380B,Bernet2012ApJ...761..144B,Hammond2012arXiv1209.1438H,Bernet2013ApJ...772L..28B,Bhat2013MNRAS.429.2469B,Joshi2013MNRAS.434.3566J,Farnes2014ApJ...795...63F,Mao2017NatAs...1..621M,Basu2018MNRAS.477.2528B}.

 For instance, the study by ~\citet[][]{Bernet2008Natur.454..302B} have used high-resolution spectra of 76 quasars and showed that quasars spectrum observations containing strong \mgii systems are associated with the larger RRM values at a wavelength of 6 cm. Subsequently, ~\citet[][]{Bernet2010ApJ...711..380B} showed that such an association of larger RRM exists only for \mgii absorbers having rest fram equivalent width (\ewr) $>$ 0.3\AA, and is absent for absorbers with   \ewr$\leq 0.3$\AA.  This was further supported by the investigation of ~\citet[][]{Bernet2013ApJ...772L..28B} where they found that sightlines with \mgii systems having impact parameter $<$ 50 kpc indeed have higher RM as compared to those with \mgii absorbers at higher impact parameter. Another technique to measure  RRM based on the radio-synthesis method has been used by ~\citet[][]{Kim2016ApJ...829..133K} which confirms that intervening systems are strongly associated with depolarization. %and characterize the complexity of the Faraday Depth spectrum using a number of parameters including intrinsic depolarization in the background sources.

Similarly, the statistical properties such as brodening of RRM distribution at different redshift bins have been used to infer the redshift evolution of cosmic magnetic field ~\citep[e.g., see][]{Welter1984ApJ...279...19W,1991MNRAS.248...58W,Kronberg2008ApJ...676...70K,2009A&A...494...21A,Bernet2012ApJ...761..144B,Hammond2012arXiv1209.1438H,2013arXiv1305.1450N,2019MNRAS.483.2424R}. For instance, ~\citet[][]{Kronberg2008ApJ...676...70K} used 268 sightlines with RM measurements at a wavelength of $\sim$10 cm in their analysis and found that the distribution of RMs' broadened with redshifts. However,~\citet[][]{Hammond2012arXiv1209.1438H}, did not find any such redshift evolution %in the width of RM distribution, where
though they used a larger sample of 3650 quasars with rotation measure compiled using a catalog of ~\citet[][]{Taylor2009ApJ...702.1230T}, based on NRAO VLA SKY SURVEY (NVSS) at a wavelength of 21 cm. This lack of firmness could be due to the fact that the sample used in this study is a mixture of sightlines with and without intervening absorbers. Since different intervening absorbers can have different absorption redshifts, they may alter their analysis of RRM pattern which is probed with various emission redshift bins of the background polarised quasars.

In an attempt to understand this discrepancy, ~\citet[][]{Bernet2012ApJ...761..144B} also analyzed the dataset of ~\citet[][]{Taylor2009ApJ...702.1230T} based on NVSS radio data at 21 cm, where they had separated the sightlines with and without intervening absorbers. They
compared their results with the work of ~\citet[][]{Kronberg2008ApJ...676...70K} and ~\citet[][]{Hammond2012arXiv1209.1438H} who too used the RM values at a wavelength of 21 cm, and with that of
~\citet[][]{Bernet2008Natur.454..302B,Bernet2010ApJ...711..380B} who used RM values at a wavelength of 6 cm.
They too did not find either an increase of the RM standard deviation with redshift or any
correlation of RM strength with \mgii absorption lines contrary to the result at 6 cm by
~\citet[][]{Bernet2008Natur.454..302B,Bernet2010ApJ...711..380B}. To reconcile this discrepancy, they have proposed an inhomogeneous Faraday screen model due to the intervening absorbers/galaxies which dilute the RRM contribution at 21 cm more as
compared to that at the wavelength of 6 cm.  Later, ~\citet[][]{Joshi2013MNRAS.434.3566J} investigated the dependence of RRM at 21 cm on intervening absorption systems with an enlarged sample consisting of 539 quasars separated out in the subsamples with and without \mgii absorbers. In their study, they found that standard deviation of RRM ($\sigma_{rrm}$) at 21 cm and the presence of \mgii absorbers are correlated though only at about 1.7$\sigma$ level, and hence at a lesser confidence level as compared to the above
studies at the wavelength of 6 cm (though using relatively smaller sample).

As an alternative route to understand the above discrepancy in the correlation of $\sigma_{rrm}$ vis-a-vis presence of \mgii absorbers at wavelength of 21 cm and 6 cm, ~\citet[][]{Farnes2014ApJ...795...63F} explored the effect of apparent frequency-dependent observational selection bias. This could arise because one may select different source populations, with different morphology and positions in relation to the optical counterparts at these high and low radio frequencies. They used the spectral index as a criterion to split their sample of 599 sources (having optical spectra, RM and spectral slope) into flat-spectrum and steep-spectrum subsamples. In their analysis, they found that the flat-spectrum subsample shows significant correlation (at $\sim$3.5$\sigma$ level) between the presence of \mgii absorption and  RM measurement at $\sim$21 cm, while that corresponding to the subsample of steep-spectrum sources do not show any  such correlation. \par 
Apart from the RRM measurement for these quasar sightlines,  their observed
fractional polarisation ($p$) is another parameter which might also be a useful tool to probe the magnetic field in the intervening galaxies, again by comparing $p$ values and its correlation with RRM for subsample{\bf s} with and without
\mgii absorbers. Many past studies have explored this possibility
and have found that indeed higher RRM is associated with the lesser $p$ value ~\citep[][]{Hammond2012arXiv1209.1438H,Bernet2013ApJ...772L..28B,Kim2016ApJ...829..133K}. For instance,
 ~\citet[][]{Hammond2012arXiv1209.1438H} found an anti-correlation between RRM and fractional polarization. One possible physical bias suggested by them was the presence of intervening absorbers, recalling the fact that RM distribution used in their study, have a mixture of sightlines with and without the \mgii absorbers.\par

It is evident from the above mentioned existing studies that while using the RM to understand the magnetic field of
the high-z galaxies and their evolution, it is important to remove various degeneracies in the observed RM values caused by different contributions  (e.g., see Eq.~\ref{rm}). This necessitates the splitting of the sample into various subsamples based on either the radio spectral slope and/or the presence/absence of the intervening absorbers. However, as pointed out above, such splitting will also lead to small
statistics problem which may not allow one to draw any firm conclusion on the nature of the magnetic field based on RM studies.
It may be noted that the main hindrance in improving the sample size in such studies is the scarcity of the
optical spectra for the sample having RM measurements, which is very crucial to separate out quasars sightlines with and without the intervening galaxies that reveal as the intervening \mgii absorbers.\par 

However, with the advent of the large spectroscopic survey such as Sloan Digital Sky Survey (SDSS) Data Release (DR)-7, 9, 12 \& 14, it has now become possible to enlarge the rotation measure sample for which optical spectra is available. At the same time, large radio catalogs of RM such as that compiled by ~\citet[][]{Taylor2009ApJ...702.1230T} consisting of  RM measurements of 37,453 sources  are now available. One also has radio spectral slope compilation such as by ~\citet[][]{Farnes2014ApJS..212...15F} consisting of 25,649 sources. It is useful to cross-correlate these optical catalogs with these large radio catalogs of RM. This forms the main motivation of our
work in this paper in which we have doubled the sample size by including the above mentioned recent SDSS Data Releases. We split the sample into subsamples based on the radio spectral slope and/or the presence/absence of the intervening absorbers. Such enlarged samples will be very useful (i) to quantify the role of intervening galaxies and their equivalent widths on RRM by comparing subsamples with and without intervening absorbers, along with their subsamples   such as steep and/or flat-spectrum radio quasars, (ii) probe any redshift evolution of RRM, and (iii) search for any  effect of the intervening absorbers on the percentage polarization of the background quasars, so as to finally infer the presence and strength of magneto-active plasma in these high redshift galaxies.

 This paper is organized as follows. We discuss in Sect. 2, the sample selection which we assembled from several optical and radio catalogs. In Sect. 3 we present the method of classification of sightlines based on detection of \mgii systems. We present our analysis and results in Sect. 4.  Discussion of our major findings and conclusions are presented in Sect. 5.
\begin{figure*}
   \epsfig{figure=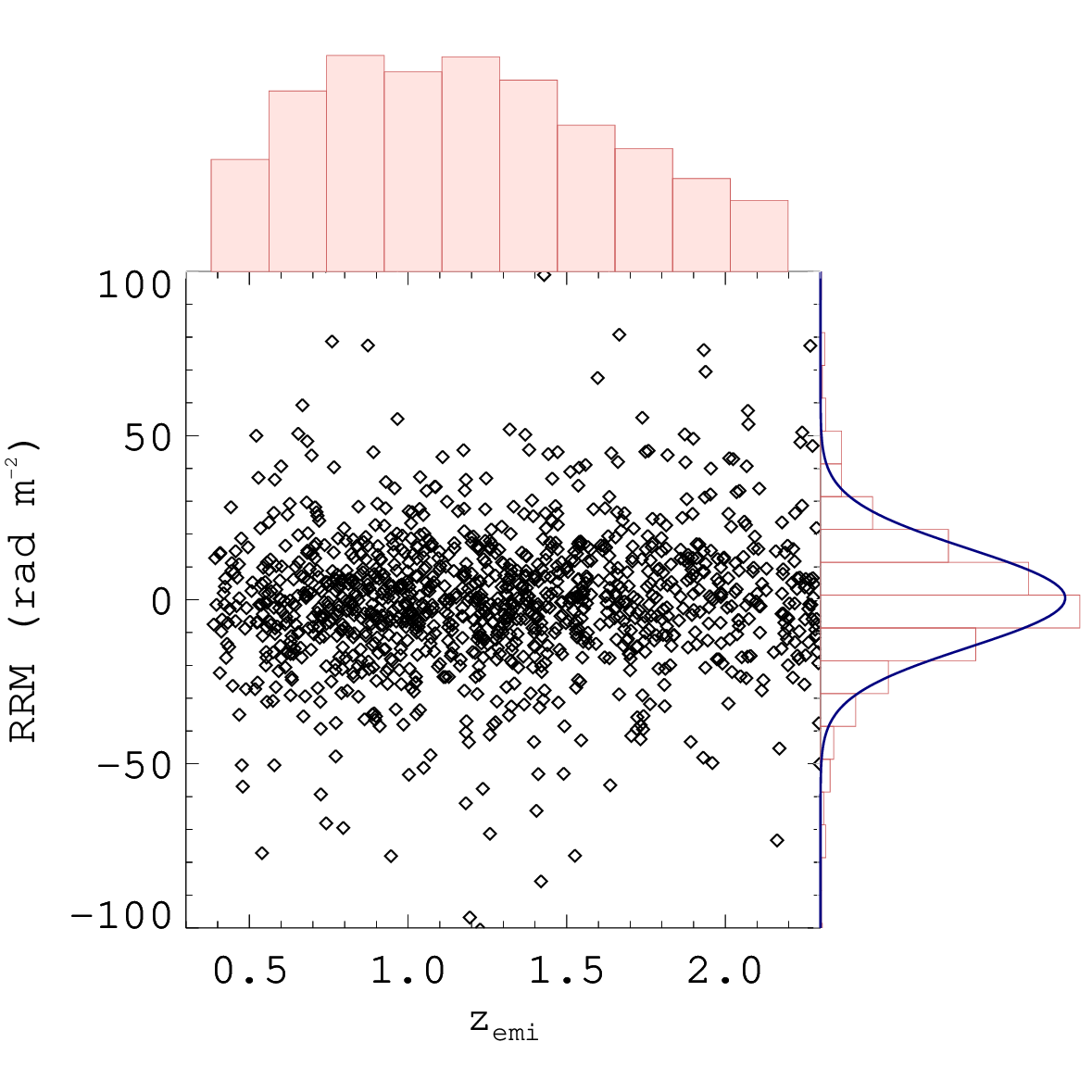,height=8cm,width=8cm,angle=0}
  \epsfig{figure=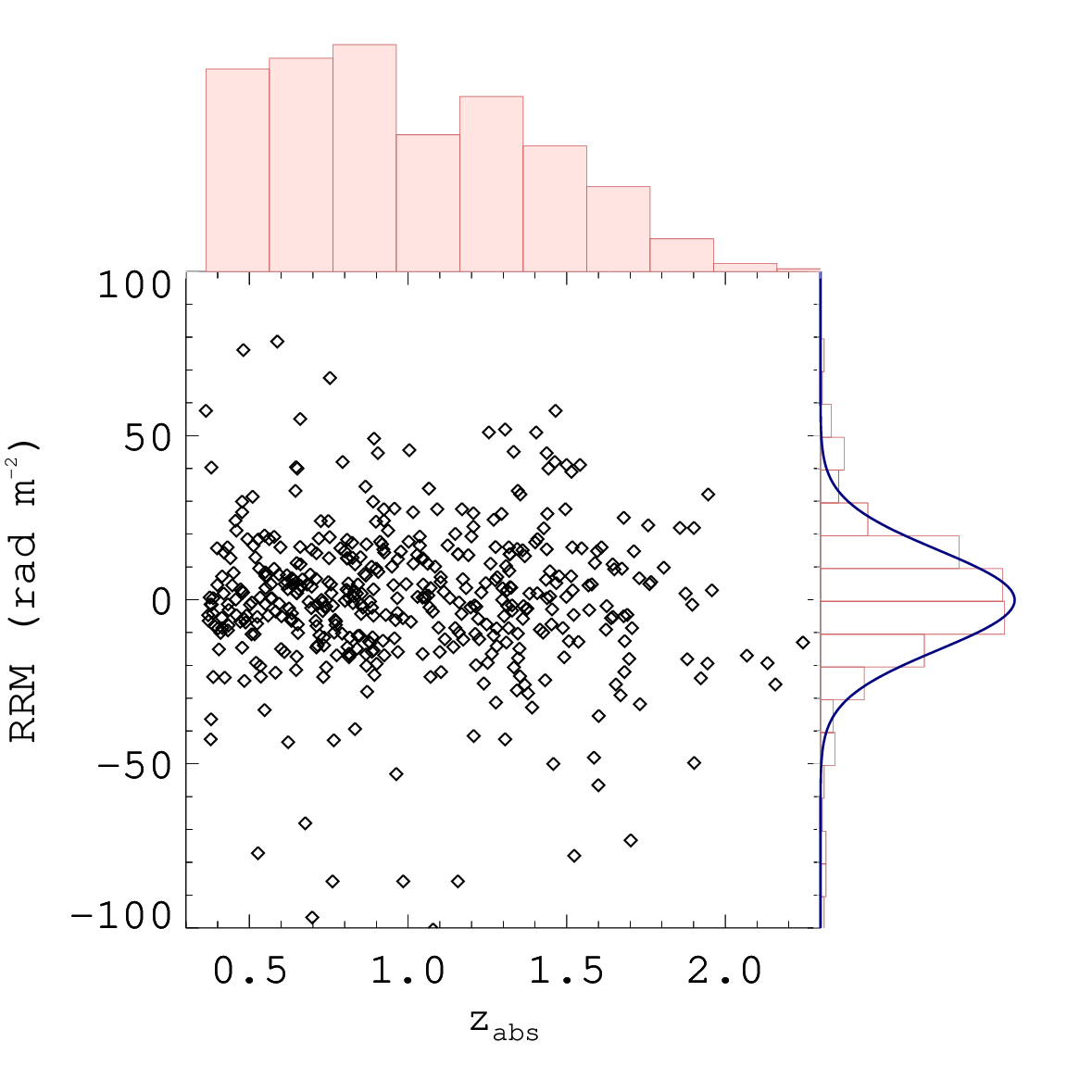,height=8cm,width=8cm,angle=0}
 
  \caption{ \emph{Left panel:} The distribution of RRM (i.e., RM$-$GRM) with the emission redshift, $z_{emi}$, for the 780 sightlines with n(\mgii) =0 along with the histogram of $z_{emi}$ on the upper axis and histogram of RRM on the right axis, fitted (using r.m.s minimization) with a Gaussian function (thick solid blue line). \emph{Right panel:} Same as left, but for 352 sightlines having n(\mgii) $>$ 0 with absorption redshift ($z_{abs}$) of \mgii absorbers.}

 \label{fig_rm}
\end{figure*} 

\section{The Sample}
\label{sample} 
 
For RM measurement of quasars, we have used a catalog of
~\citet[][henceforth TSS09]{Taylor2009ApJ...702.1230T}\footnote{http://www.ucalgary.ca/ras/rmcatalogue} 
consisting of RM values for 37,543 sources based on NVSS observations. In this they have used Q and U Stokes Parameters in two frequency bands, 1.36 GHz and 1.43 GHz to estimate the polarised intensity and percentage polarisation.

For searching the optical counterpart of the quasars listed in TSS09, we have used SDSS DR-7, 9, 12 \& 14 catalogs of quasars for their optical spectra by putting constrain on  emission redshift ($z_{emi}$) of sources to be in a  range $0.38 \le z_{emi} \le 2.3$. Here, the upper redshift cutoff of 2.3 is imposed based on the upper limit of 9000 \AA\ on the wavelength of the SDSS spectrum. For quasars with redshift more than this limit, the \mgii emission lines is beyond the SDSS spectral coverage. Hence, our upper limit avoids the ambiguity  of any \mgii absorber falling above the spectral coverage. Our lower limit of 0.38 on the emission redshift ensures that the starting observed wavelength
 of SDSS spectra at 3800 \AA~allows us to detect at least one \mgii doublet, if present. Further, we cross-correlate the RM catalog of TSS09 with all the quasars (i.e., from the SDSS DR-7, 9, 12 \& 14 catalogs) fulfilling the above redshift criterion (i.e., $0.38 \le z_{emi} \le 2.3$), by demanding that their optical and radio positions match within an offset of less than 7 arcsec. This optimal offset of the 7 arcsec was chosen based on the analysis of ~\citet[][]{2018MNRAS.480.1796S}, where they showed that at NVSS resolution this degree of tolerance is optimal in such a cross-correlation with optical positions of SDSS catalogs.\par
 The details of the outcome of our above cross-correlation among these catalogs are given in Table~\ref{table:sample}.

\begin{table}
\begin{center}
%\begin{minipage}[10]{140mm}
\caption{Details about our selection of 1135$^d$ quasars sample.}
\label{table:sample}

\begin{tabular}{@{}p{2cm} p{1.3cm} p{1.3cm} p{1.3cm} p{1.3cm}@{}} 
\hline \hline

%\multicolumn{2}{c|}{Optical catalog of \mgii system}  &  %\multicolumn{2}{c}{Cross-match with TSS09 RM catalog} &  \\
\hline
Criteria     & DR7 & DR9  & DR12 & DR14 \\
\hline
Total       & 84533 & 15439& 297301 & 525982\\
z-range$^{a}$   & 75450 & 7640 & 148871 & 285042 \\
TSS09-match$^{b}$ & 673 & 51 & 404& 588\\
Taken$^{c}$       & 673 & 23 & 257& 182\\

\hline   \\
\multicolumn{5}{l}{$^{a}$ $0.38 \leq z_{emi} \leq 2.3$, based on SDSS spectral coverage;}\\
\multicolumn{5}{l}{$^{b}$Number of sources found common  within 7 arcsec between}\\
\multicolumn{5}{l}{various SDSS data release and the TSS09 RM catalog consisting } \\
\multicolumn{5}{l}{RM values for 37,543 sources; $^{c}$The number of sources taken} \\
\multicolumn{5}{l}{from the cross-match found for our sample, after removal of any }\\
\multicolumn{5}{l}{repeated sources among the above SDSS data release.}\\
\multicolumn{5}{l}{$^d$ Also included 3 outlier $>9\sigma$, our final sample consist of}\\
\multicolumn{5}{l}{1132 sources (e.g., Table ~\ref{table:sample1})}
%\multicolumn{5}{l}{}\\
\end{tabular}     
%\end{minipage}     
 \end{center}                                                     
\end{table}
As can be seen from this table (last row), after removal of the quasar duplications among various data releases of SDSS, the SDSS DR-7, 9, 12 and 14 have contributed 673, 23, 257 and 182 quasars respectively, leading to the merged sample of 1135 sources. In fact, we also note here that ~\citet[][]{Xu2014MNRAS.442.3329X}  have also complied RM data for 4553 sources using observations in the frequency range from 1.3 GHz to 2.3 GHz. However, to avoid the inhomogeneity due to the mixing of two catalog, we have only used the (larger) catalog TSS09 as our reference catalog for RM measurement. By  not using the sample of~\citet[][]{Xu2014MNRAS.442.3329X}, only  a nominal decrements of about  2.7\%  occurs in  our final sample size.

Further, we have noted that in the catalog of TSS09 based on NVSS the value of GRM is not available. Hence, we have made use of the GRM compilation by  ~\citet[][]{Xu2014MNRAS.442.3329X} based on their online GRM calculator\footnote{http://zmtt.bao.ac.cn/RM/searchGRM.html}. Out of the total of 1135 sources selected by us, 134 sources were common with ~\citet[][]{Xu2014MNRAS.442.3329X} and hence we could use the GRM values for them directly from this catalog. For the remaining 1001 sources, the  above  online GRM calculator has been used to estimate their GRM values. Here, we have assumed that the error in GRM due to the difference in the measurement frequency (1.3 to 2.3 GHz in the  ~\citet[][]{Xu2014MNRAS.442.3329X} and 1.4 GHz in the TSS09) is negligible as compared to the typical uncertainty in the GRM measurements themselves. \par

The fraction of our sources with $\vert RRM \vert$ values $\le$ 25, 50 and 100 ~\rrm~is typically found to be 81.89\%, 96.11\%, and 99.55\%, respectively. In this distribution, we noted three outliers in RRM with values of 142.4 (SDSS J012142+114950), $-$473.8 (SDSS J111857+123442) and $-$172.1 \rrm~(SDSS J110120+415308) which deviate from the mean at more than $9\sigma$~level. In order to avoid the dominance of these outliers especially on the mean and standard deviation we have excluding them from our analysis and hence are left with 1132 sources for our final analysis, with their details as given in Table~\ref{table:sample1}.

%%%%%%%%%%%%%%%%%%%%%%%%%%%% SAMPLE %%%%%%%%%%%%%%%%%%%%%%%%%%%%%%%%%%%%%%%%%%%%
%%%%%%%%%%%%%%%%%%%%%%%%%%%%%%%%% Table1%%%%%%%%%%%%%%%%%%%%%%%
\begin{table*}
\centering
\caption{Main properties of our 1132 quasars sample (after discounting 3 outliers being at $9\sigma$ level).}
\label{table:sample1}
{\scriptsize
 
\begin{tabular}{c r r r  p{1cm} r  r c p{1cm}p{1cm}p{1cm}p{2cm}}
\hline \hline
SDSS Name     & RM  & $\delta RM$ & GRM    & $\delta GRM$   & RRM    & $\delta RRM$  &  n(Mg~{\sc ii})   &$z_{emi}$ &  $p$& $\delta p$ & $\alpha^{a}$\\
              &   \multicolumn{6}{c}{\rrm}                                           &             &          & (\%)&  (\%)       &  \\
(1)           &(2)  & (3)         & (4)    & (5)            & (6)    & (7)           & (8)         & (9)      & (10)& (11)        & (12)  \\

\hline
J142746+002848    &    25.3   &     13.6  &       3.9  &       6.1   &     21.4  &      14.9 &    0  & 1.2628  &  5.97 & 0.3 & -~-~-~-    \\
J003936+204912    & $-$23.3   &      6.5  &   $-$24.4  &       8.2   &      1.1  &      10.5 &    0  & 1.3778  &  3.07 & 0.1 &$-$0.77  \\
J095443+403636    &  $-$3.1   &      3.8  &    $-$1.0  &       4.2   &     $-$2.1&       5.6 &    1  & 0.7821  &  5.53 & 0.1 &$-$0.97  \\
J083824+123000    &    30.8   &      9.4  &      29.8  &       7.3   &      1.0  &      11.9 &    2  & 1.6221  &  3.75 & 0.2 &$-$0.96  \\
...               &  ...       & ...        &     ...     &     ...      &       ...  &       ...  &    ...&    ...&  ...  &   ...& ...    \\

\hline
\multicolumn{12}{l}{Note: The entire table is available in the online version. Only a portion of this table is shown here, to display its form and content.}\\
\multicolumn{12}{l}{$^a$ Radio spectral indices, $\alpha$ define by $F_{\nu}\propto \nu^{\alpha}$ has been taken from ~\citet[][]{Farnes2014ApJS..212...15F}.} 

\end{tabular}     
}
\end{table*}

\begin{figure}
  
  \epsfig{figure=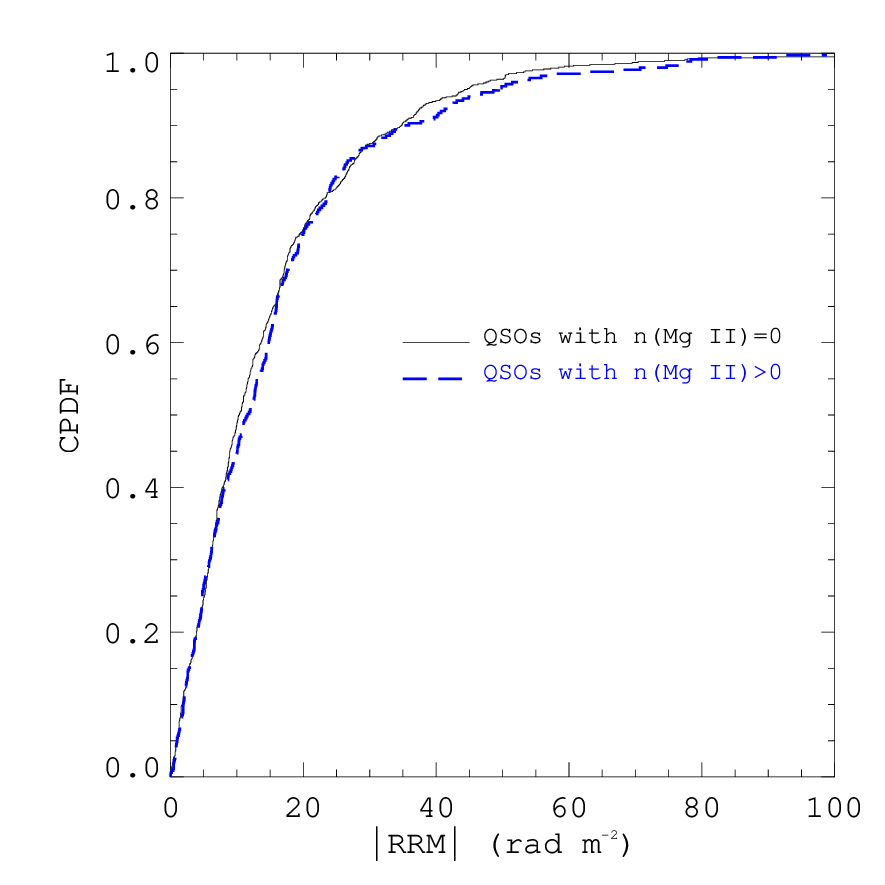,height=8cm,width=8cm,angle=0}

\caption{Cumulative probability distribution (CPDF) of the absolute value of RRM (i.e., $|RRM|$), for the quasar sightlines with (blue dashed line), without \mgii absorbers (black solid line).} 
\label{fig:cuml_rrm_zgt1}
\end{figure}

%%%%%%%%%%%%%%%%%%%%%%%%%%%% SAMPLE %%%%%%%%%%%%%%%%%%%%%%%%%%%%%%%%%%%%%%%%%%%%
%%%%%%%%%%%%%%%%%%%%%%%%%%%%%%%%% Table1%%%%%%%%%%%%%%%%%%%%%%%
\begin{table}
\centering

%\begin{fullpage}
\caption{Main properties of our 442 \mgii absorption systems seen towards 352 quasars sightlines in our sample.}
\label{table:sample2}
{\scriptsize
\begin{tabular}{@{}cp{1cm}p{1cm} cc@{}} 
\hline \hline
SDSS Name   & $z_{emi}$  &  $z_{abs}$ & $EW_{r}$(\AA) &  $\delta EW_{r}$(\AA) \\
%\multicolumn{3}{c|}{Optical catalog of \mgii system}  &  \multicolumn{2}{c}{Cross-match with NVSS catalog} &  \multicolumn{2}{c}{} \\
(1)              &(2)            &(3)               & (4)      & (5)                  \\
\hline
J095443+403645   &         0.7821 &        0.742    &    0.63  &       0.09 \\
J024534+010814   &         1.5296 &        1.113    &    1.77  &       0.17 \\
J003032$-$021156 &         1.8048 &        1.372    &    1.71  &       0.08 \\
J003032$-$021156 &         1.8048 &        1.454    &    1.25  &       0.10 \\
...              &        ...   &     ...        &     ...     &   ...      \\
...              &        ...   &     ...        &     ...     &   ...      \\ 
\hline
    
\multicolumn{5}{c}{Note: The entire table is available in the online version. Only a }\\
\multicolumn{5}{c}{portion of this table is shown here, to display its form and content.}\\

\end{tabular}     
}

\end{table}

 \begin{figure}
   
   \epsfig{figure=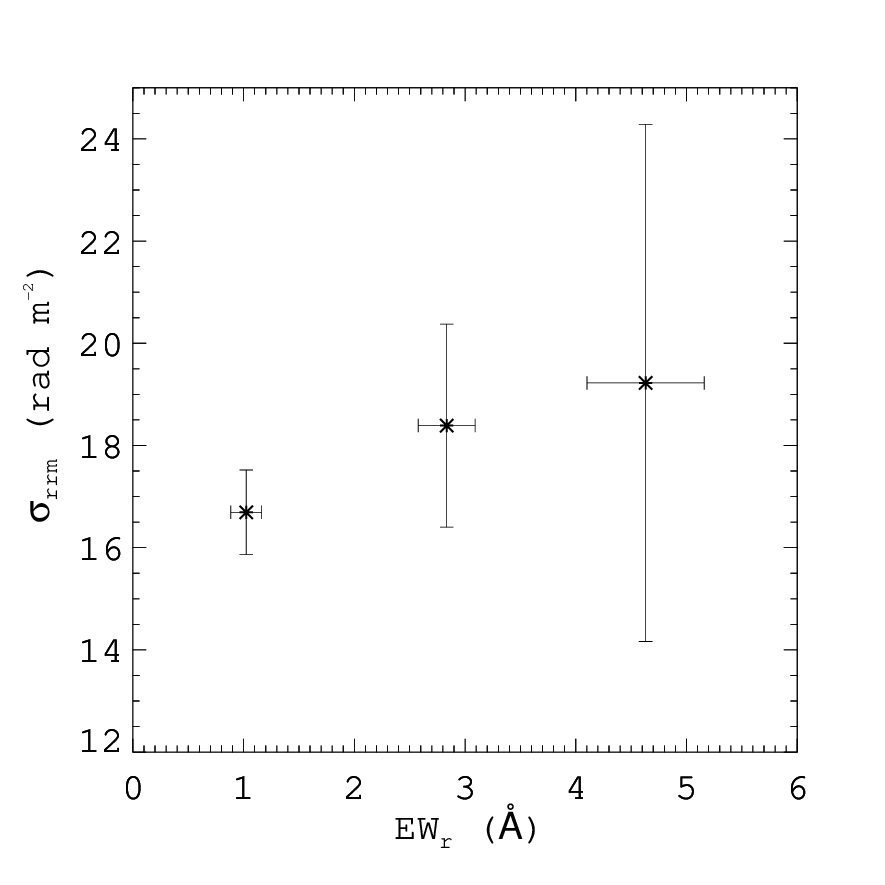,height=8.cm,width=8.cm,angle=0}
   \caption{The plot of RRM broadening using MADFM (i.e., $\sigma^{md}_{rrm}$) in different \ewr\ bin, based on 352 quasars with \mgii intervening absorbers. Here for sightline with multiple absorber, the \ewr\ of such individual sightline has been summed up.}
  
  \label{fig:rrm_vs_ew1}
   \end{figure}
\section{Identification of \mgii systems}
\label{identification}
The identification of the \mgii absorption doublet in the normalized continuum spectrum was carried out using the procedure discussed in detail by ~\citet[][]{Joshi2013MNRAS.434.3566J}~\citep[see also,][]{sapna2018}. Briefly, the procedure automatically searches for absorption features in the normalized spectrum in the range $(1+z_{emi})\times 1216 \textless \lambda \textless (1+z_{emi})\times 2803$ \AA. The search was carried out of such individual sightline for absorption features, fitted with a Gaussian profile by taking an initial full width at half-maximum (FWHM) of 2.5 pixels. Out of all probable cases, the  selection was made by accepting only the lines with line depth (i.e., at centroid) above three times the error-bar in that region of the spectrum. The absorption features thus identified for each quasar were searched for absorption line pairs. For this purpose, in our procedure we first computed the redshift of a given absorption feature, assuming it to be \mgiia. The corresponding positions of the expected \mgiib~ and \feiia~ lines were then inspected visually in their velocity plots. Here, we have also used the line profile matching technique of doublet.  For this, we first plotted the normalized spectrum around the candidate \mgii doublet and then overplotted the same spectrum by shifting the wavelength axis by a factor of Mg~{\sc ii}$\lambda 2796.3543$/Mg~{\sc ii}$\lambda 2803.5315$ (i.e., 0.997). The location of reasonable overlap between the absorption lines in the shifted and the original (unshifted) spectra were marked as a detected \mgii absorption system.

For each detected \mgii absorption system, we also assessed the quality of our  underlying continuum fit during our 
visual  check. If deemed desirable we carried out a local continuum fitting and this improved fit was then used to obtain a better estimate of equivalent width of \mgii lines. Out of total 1132 sightlines in our sample we found that 352 quasars sightlines have at least one \mgii absorber and the remaining 780 quasar sightlines are without any such \mgii absorbers. Among the 352 sightlines, 278 have one absorber each, 63 have two absorbers each and 11 have more than two absorbers, resulting in a total detection of 442 \mgii absorption systems, as listed in Table~\ref{table:sample2} along with details of their $z_{emi}$, $z_{abs}$ \& \ewr. In Fig.~\ref{fig_rm} we show the plot of RRM versus $z_{emi}$ for sightlines without intervening absorbers, (i.e., n(Mg~{\sc ii})$=$0) and RRM versus $z_{abs}$ for sightlines with n(Mg~{\sc ii})$>$0 along with their histograms (e.g., see right and upper side of plots). As can be noted from the RRM histogram plot, a Gaussian fit (using r.m.s minimization) describes RRM distribution reasonably well both for subsample with and without \mgii absorbers. We will be using this aspect while computing the difference of standard deviation (in quadrature e.g., see Eq. \ref{eq:std_excess}) among the sightlines with and without \mgii systems. \par

We also noted here that the compilation of \mgii systems for all the quasars belonging to SDSS DR$-$7, 9 is also reported by ~\citet[][henceforth ZM13]{Zhu2013ApJ...770..130Z}, for DR$-$12 by using their online catalog\footnote{https://www.guangtunbenzhu.com/jhu-sdss-metal-absorber-catalog} (henceforth ZM17) and for SDSS DR$-$14 by ~\citet[][henceforth RS16]{Srinivasan2016MNRAS.463.2640R}. However, for the sake of uniformity in identifying the \mgii system among the members of our sample belonging to the various above mentioned SDSS data releases, and also due to the importance of the visual check (e.g., for profile matching and continuum errors) as outlined above, our analysis relies on our own \mgii system identification. This is particularly important as lack of visual check might not have effect on  analysis using few thousands of sightline (e.g., in ZM13, ZM17 and RS16), but may have significant impact in the analysis like in this study where only a few hundred quasars contribute in a subsample.  Just for comparison we noticed that the  ZM13, ZM17 and RS16 catalogue (based on automated search) missed genuine  27, 31 and 17 \mgii systems, respectively. On the other hand the number of false detection found were 10, 5 and 2 \mgii systems, respectively. The difference is only about $\sim$8\%.  However, it can still significantly affect the RRM analysis which also illustrates  the importance of supplementing the  automated \mgii search by the visual check as adopted in our procedure of \mgii identification.

\section{Analysis and Results}
\subsection{Comparison of RRM distribution for subsample with and without \mgii absorbers}
\label{result1}

 To carry out the comparison of the RRM distribution for subsamples with and without \mgii absorbers, we have shown the cumulative probability distribution (CPDF) of their absolute value of RRM, ($\vert RRM \vert$) in Fig.~\ref{fig:cuml_rrm_zgt1}. Usually Kolmogorov-Smirnov test (KS-test) is used in the literature to quantify any statistical difference between two such distributions. However, for distributions which differ mainly in their tailed regions, it is found that the Anderson-Darling\footnote{\hyperref[]{https://asaip.psu.edu/Articles/beware-the-kolmogorov-smirnov-test}} test (hereafter, AD-test) could be a preferred test instead of KS-test.
  % give better result when two distributions are showing more difference in tails of the distributions.
  Therefore we use the AD$-$test throughout except while  comparing with the results in the literature based on the KS$-$test. For the above $\vert RRM \vert$ distributions in the subsamples with and without \mgii absorbers, the null hypothesis probability, that 
  these two distributions are drawn from the same distribution,  is
  found to be $\sim$68\%. This shows that the functional form of their CPDF is almost same. However, it can be noticed from the CPDF plots shown in the Fig.~\ref{fig:cuml_rrm_zgt1} that these two CPDFs cross each other multiple times in the $\vert RRM \vert$ range between 15 to 60, being smoother for the case of the subsample with n(Mg~{\sc ii})$=$0. Such $\vert RRM \vert$ fluctuations in the sample with n(Mg~{\sc ii})$>$0 could also be due to the probable scatter caused by the presence of magnetic filed in the intervening galaxies responsible for the \mgii absorbers. This is over and above a  possible smearing  out of any difference due to the range in redshift and radio spectral indices (Sects. ~\ref{result2} \& ~\ref{red_evol}). Therefore RRM distribution being an important observational parameter, we have quantified it by measuring its broadening in subsample with \mgii absorbers in comparison to the subsample without such absorbers (even though their CPDF form may be similar apart from scatter).  

  For this commonly used method is to compute the standard deviation (SD) of the subsample as,
\begin{equation}
%\scripsize
  \sigma_{rrm}^{sd} = \sqrt{\frac{\sum_{i=1}^{N} (x_{i}-\bar{x})^2 }{(N-1)}},
    \label{sigmasd}
\end{equation}

with error ($\delta \sigma_{rrm}^{sd}$) given by,

\begin{equation}
%\scripsize
  \delta \sigma_{rrm}^{sd} = \frac{\sqrt{\sum_{i=1}^{N} (x_{i}-\bar{x})^2 (\delta x_{i})^2 + (\sum_{i=1}^{N} (x_{i}-\bar{x}))^2 (\delta \bar{x})^2}}{(N-1)\times \sigma_{rrm}^{sd}}.
    \label{err_sd}
\end{equation}

Here, $x_{i}$ and $\delta x_{i}$ is the RRM measurements and error on the RRM, $\bar{x}$ and $\delta \bar{x}$ represents the mean and the error on the mean of the RRM measurements, respectively. We get the $\sigma_{rrm}^{sd}$ for the subsample of with and without intervening \mgii systems, $22.9\pm0.6$ ~\rrm~ and $21.6\pm0.4$~\rrm, respectively, as listed in Table~\ref{table:sigma}. \par
We also note that such a comparison, quantified by $\sigma_{rrm}^{sd}$, might also get dominated by the presence of few outliers of the samples. A better alternative is to employ the median absolute deviation from the mean, MADFM (MD hereafter) rather than $\sigma_{rrm}^{sd}$, the former being also resistant to outliers and is computed as:

\begin{equation}
%\scripsize
  MADFM = Median(\vert x_{i} - \bar x\vert)
 \label{mad}
\end{equation}

The MADFM can be used as a consistent estimator for the estimation of spread, analogous to $\sigma_{rrm}^{sd}$, as $\sigma_{rrm}^{md} = k\times$ MADFM with $k =1.4826$ for normally distributed data ~\citep[][]{article}. This  is a reasonable approximation for our  RRM distribution as revealed by the Gaussian fit of RRM shown in  Fig.~\ref{fig_rm}, apart from the minor deviation due to \mgii systems we are searching for in sample with \mgii absorber vis-a-vis without such absorbers. Therefore, we will be quantifying the broadening of our RRM distribution using the $\sigma_{rrm}^{md}$
 unless otherwise specified (e.g., when comparing with literature results based on the SD-method).

We found that $\sigma_{rrm}^{md}$ for subsample with intervening absorbers (i.e. with n(Mg~{\sc ii})$>$0)  and without these absorbers (i.e. with n(Mg~{\sc ii})=0) to be $17.1\pm0.8$~\rrm\ and 15.0$\pm0.6$~\rrm,\ respectively (see Table~\ref{table:sigma}). Here, the error bar on the MADFM value is  calculated by using a standard procedure applicable for the error on a mean value of such deviations (i.e., $\langle \vert x_{i} - \bar x\vert \rangle$). %x_i-\bar x \rangle$) 

To quantify the excess broadening ($\sigma^{ex}_{rrm}$)
introduced by the intervening absorbers, we have
subtracted in quadrature the $\sigma_{rrm}$ (assuming Gaussian distribution, e.g., as shown in  Fig.~\ref{fig_rm}) for the quasar subset with ($\sigma_{rrm(w)}$) and without ($\sigma_{rrm(wo)}$) \mgii absorber as,
%%%%%%%%%%%%%%%%%%%%%%%%%%%% Sigma %%%%%%%%%%%%%%%%%%%%%%%%%%%%%%%%%%%%%%%%%%%%

%%%%%%%%%%%%%%%%%%%%%%%%%%%% Sigma %%%%%%%%%%%%%%%%%%%%%%%%%%%%%%%%%%%%%%%%%%%%
%%%%%%%%%%%%%%%%%%%%%%%%%%%%%%%%% Table2%%%%%%%%%%%%%%%%%%%%%%%
\begin{table*}
 \begin{center}
%\begin{minipage}[10]{180mm}
\caption{Results of RRM distribution for various subsample.} 
%\HC{This form may be compact. Footnote are commented here, SM may add later beside filling the xx}}
\label{table:sigma}
\begin{tabular}{@{}l r r r r r r r c r @{}}
  \hline \hline

\multicolumn{1}{c}{Sample type}
&\multicolumn{3}{c}{n(Mg~{\sc ii}) = 0}
&\multicolumn{3}{c}{n(Mg~{\sc ii}) $>$  0}
&\multicolumn{2}{c}{$\sigma^{ex}_{rrm}$ using MADFM/SD}
&\multicolumn{1}{c}{$P_{null}$(\%)}\\
\hline
% --next line of unit
&\multicolumn{1}{c}{N}
&\multicolumn{1}{c}{$\sigma_{rrm}^{md}$}
&\multicolumn{1}{c}{$\sigma_{rrm}^{sd}$}
&\multicolumn{1}{c}{N}
&\multicolumn{1}{c}{$\sigma_{rrm}^{md}$} %(see Eq. \~\ref{mad}).}
&\multicolumn{1}{c}{$\sigma_{rrm}^{sd}$}
&\multicolumn{1}{c}{$\sigma_{md}^{ex}(n\sigma)$}
&\multicolumn{1}{c}{$\sigma_{sd}^{ex}(n\sigma)$}
&\multicolumn{1}{c}{AD$^{a}$} \\

\hline 

        \hline
        \\
        %\multicolumn{10}{c}{Full sample statistics.}\\
        %\\
        \hline
	    Full   & $780$   & $15.1\pm0.6$ & $21.6\pm0.4$ & $352$ &  $17.1\pm0.8$ & $22.9\pm0.5$ &   $8.0\pm1.9 (4.2\sigma)$ & $7.5\pm 2.0(3.7\sigma)$ & 68               \\ 
%\hline
%\\
%\multicolumn{10}{c}{Statistics based on radio spectral index ($\alpha$); $F_{\nu} \propto \nu^{\alpha}$.}\\ 
%\\
%\hline
		$\alpha$ $\ge$ $-$ $0.3^{b}$    & $204$ & $14.3\pm1.1$ &  $22.3\pm0.8$ & $111$  & $17.3\pm1.3$ & $22.0\pm1.1$ & $9.8\pm2.8$(3.5$\sigma$)   & -~-~-~-  & $24$  \\  
		%3.70$\pm7.90^{g}$
        $\alpha$ $\le$ $-$ $0.7^{c}$    & $336$ & $15.2\pm0.9$ &  $21.7\pm0.6$ & $140$  & $15.6\pm1.2$ & $19.4\pm0.9$ & $3.4\pm7.2$(0.5$\sigma$)  & -~-~-~- &  69 \\ 
        %9.65$\pm2.29^{g}$
\hline
\\

        \multicolumn{10}{l}{$^{a}$The percentage probability of null hypothesis using Anderson-Darling(AD) test for a subsample with respect to n(Mg~{\sc ii}) = 0 }\\
        \multicolumn{10}{l}{subsample. $^{b}$ Flat spectrum radio quasars; $^{c}$ Steep spectrum radio quasars.}

\end{tabular}
                                                          
%\end{minipage}      
\end{center}                                                     
\end{table*}
%%%%%%%%%%%%%%%%%%%%%%%%%%%%%%%%%%%%%%%%%%%%%%%%%%%%%%

\begin{equation}
\sigma^{ex}_{rrm} =  \sqrt {\sigma_{rrm(w)}^2 -  \sigma_{rrm(wo)}^2 }.  
\label{eq:std_excess}
\end{equation}
and its associated  error as,

\begin{equation}
%\scripsize
\delta \sigma^{ex}_{rrm} = \frac{1}{\sigma^{ex}_{rrm}} \sqrt {
  \sigma_{{rrm (w)}}^2 \delta \sigma_{rrm (w)}^2 +
  \sigma_{rrm (wo)}^2 \delta \sigma_{rrm (wo)}^2 }.  
\label{eq:std_excess_pro}
\end{equation}

This excess is found to be 8.0$\pm1.9$~\rrm~(i.e., at 4.2$\sigma$ level) and 7.5$\pm 2.0$ ~\rrm~(i.e., at 3.7$\sigma$ level) while estimating the broadening of the RRM using MADFM (i.e., $\sigma_{rrm}^{md}$) and  SD (i.e., $\sigma_{rrm}^{sd}$) method, respectively, as listed in Table~\ref{table:sigma}.\par

We note here that, the results derived to quantify the difference between the subsample with and without \mgii absorbers, by comparing the $\sigma$'s of their RRM  distributions does show discrepancy vis-a-vis derived based on the estimation of  null probability (using AD-test). Such discrepancy is unlikely if the true distribution of the RRM is Gaussian as we have assumed in our estimation of $\sigma^{ex}_{rrm}$ based on Eq.~\ref{eq:std_excess}. One possibility to understand this discrepancy is to use both these tests on  simulated datasets of RRM having Gaussian distributions. The  simulated datasets for subsamples with and without \mgii systems are different in their standard deviation which is taken from their observed values of $15.1\pm0.6$ and $17.1\pm0.7$ \rrm, respectively. We have simulated 1000 realisations of these Gaussian random distributions for the subsamples with (352 sightlines) as well as without (780 sightlines) \mgii intervening absorbers.  \par

We have evaluated the $P_{null}$  based on AD-test for all these 1000 simulated datasets of RRM, among  with and without \mgii systems, and found that for most of the realisations ($\sim 70\%$) it is less than $10\%$, which is much less than the $P_{null}$ of $68\%$ found among real data set.

This suggests that the RRM distributions in real datasets  perhaps is not strictly Gaussian, which may be a reason for the above observed discrepancy between the results based on $\sigma^{ex}_{rrm}$ (assuming Gaussianity of RRM distribution) and the $P_{null}$ based on the AD-test. However, in a larger perspective and to compare our results with earlier similar studies (e.g., see ~\citet[][]{Joshi2013MNRAS.434.3566J} and ~\citet[][]{Farnes2014ApJ...795...63F}) which has  assumed Gaussian distribution (hence made use of Eq.~\ref{eq:std_excess}), we here retain our result based on this technique/estimator to spot the effect of the magnetised plasma in these absorbers on the rotation measure of the background quasars.

Furthermore the simulated datasets can also be used to quantify any chance probability of obtaining $\sigma^{ex}_{rrm}=8\pm 2$~\rrm (i.e., at $4\sigma$ level), as we have explored here only limited parameter space and the effect such as look-elsewhere effect ~\citep[][]{look2010} may also be at play. To explore any  such possibility we have carried out Monte-Carlo simulations (for $10^{6}$ random realisations) of our subsample with n(Mg~{\sc ii})=0 by allowing the RRM values  to vary within its  $1\sigma$ errorbar. Then based on the distribution of the $\sigma^{ex}_{rrm}$  values (as in Eq.~\ref{eq:std_excess}) between the median value of the $\sigma$ for all $10^{6}$ realisations and the values of sigma corresponding to each of the random subsample of 352 RRM values from each realisation, we noticed that the observed excess of $\geq8$~\rrm~ (as we found in our analysis) has its by chance probability about $\sim18\%$. This suggest that the typical confidence level of the $\sigma^{ex}_{rrm}$  could be 1-2$\sigma$ rather than the $\sim 4\sigma $ (based on the observed $8\pm2$~\rrm).
\par

Further, to check the impact of the strength of the absorber on RRM distributions using our 352 sightlines with intervening absorbers, we also plotted the  $\sigma^{sd}_{rrm}$ at different \ewr\ bin as shown in Fig.~\ref{fig:rrm_vs_ew1}. For this purpose we have used the summed \ewr\ for sightlines with multiple absorbers. 

Here we did not find any firm trend between $\sigma^{md}_{rrm}$ with \ewr, but statistics with large number of observations of such systems can do better clarification to this correlation. 

It is important to use the above estimate of excess in $\sigma^{md}_{rrm}$ 
of subsample with \mgii systems as compared to the subsample without such systems to get a typical estimate of the average strength of the parallel component of magnetic fields ($\langle B_\parallel\rangle$) (rest frame) in these high-z galaxies responsible for the \mgii absorption systems. For this we have used the formalism similar to the~\citet[][]{Kronberg2008ApJ...676...70K} (e.g., see their Eq. 15) as;

\begin{eqnarray} \nonumber
\label{eq:magne} \langle B_\parallel\rangle &=& 5.5\times10^{-7}  
{\rm G}
\left( \frac{1 +\langle z_{abs}\rangle}{3.5} \right)^{ 2} \\ & \times &
\left(\frac{\sigma^{ex}_{rrm}}{20\;{\rm rad~m}^{-2}}\right) \;
\left(\frac{N_e}{1.7\times 10^{21}{\rm cm}^{-2}}\right)^{-1}
\label{mag}
\end{eqnarray}

Here, $N_{e}$ is the  column density of electrons in the intervening absorber systems  with a typical value of $N_{e} \sim 9\times10^{19} cm^{-2}$ ~\citep[e.g., see,][]{Bernet2008Natur.454..302B}, and $\langle z_{abs}\rangle$ is the median redshift of the absorbers.
%, whereas $\sigma^{ex}_{rrm}$ is the excess in the $\sigma_{rrm}$ in each subsample.
Using, $\sigma^{ex}_{rrm}=8.0\pm 1.9$~\rrm~ (using MADFM) obtained for our full sample with and without \mgii systems (e.g., see Table~\ref{table:sigma}, top row), the typical value of  $\langle B_\parallel\rangle$ is found to be around $\sim 1.3\pm 0.3~\mu G$, in these intervening galaxies spread over the redshift range from 0.38 upto 2.24 with median redshift of 0.92.

\subsection{Correlation of RRM  using subsample of  flat and steep spectrum sources}
\label{result2}
\begin{figure}
 
   \epsfig{figure=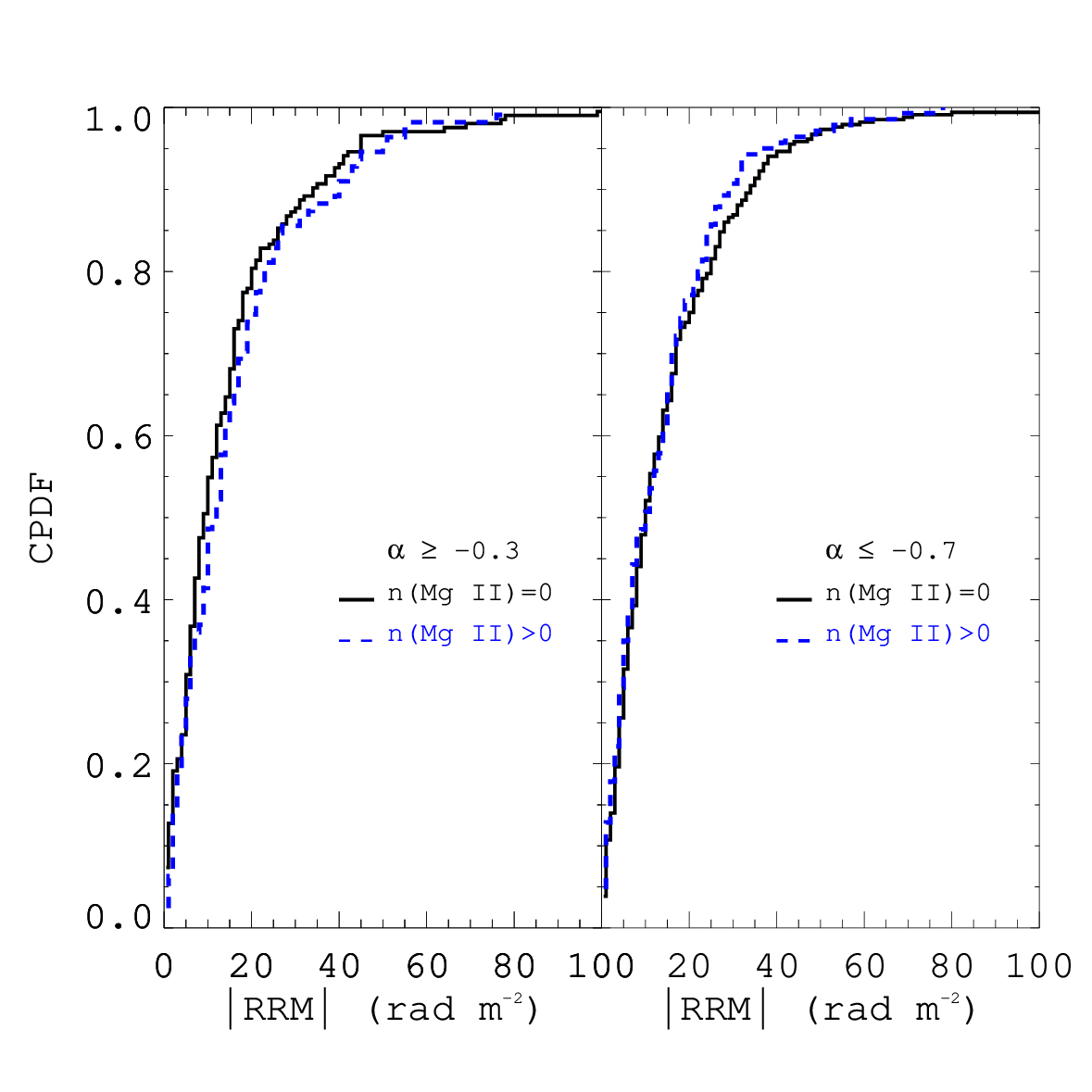,height=8.0cm,width=9.0cm,angle=0}

   \caption{\emph{Left panel:} The CPDF of $|RRM|$ for the flat spectrum quasar with $\alpha \geq -0.3$ (315 sources) for their subsamples without (black solid line) and with \mgii absorbers (blue dashed line). \emph{Right panel:} Same as left, but for steep spectrum quasars with $\alpha \le -0.7$ (476 sources).}

\label{aplha_cuml_rrm1}
\end{figure}

As pointed out by ~\citet[][]{Farnes2014ApJ...795...63F}, the morphology of radio sources can also introduce a bias in the correlation of RRM distribution of sightlines with \mgii absorbers, such as, compact flat spectrum sources being more aligned with the optical sightlines in comparison to the possible misalignment of lobes dominated steep spectrum sources. In their study, they found a significant correlation (at $\approx 3.5\sigma$ level) between the \mgii absorbers and RM in the subsample of flat spectrum (core dominated, with $\alpha$ $\geq -0.3$; for $F_{\nu} \propto \nu^{\alpha}$) sources. Such a correlation is, however, absent in the subsample of steep spectrum (lobes dominated, with $\alpha$ $\leq -0.7$) sources. We have also carried out an analysis similar to that in 
~\citet[][]{Farnes2014ApJ...795...63F} by making use of our bigger sample of 1132 sources. For the values of spectral indices, we make use of~\citet[][]{Farnes2014ApJS..212...15F} compilation 
consisting of 25649 sources. After cross-correlating their sample with
our sample of 1132 sources, we could get the $\alpha$ values for 1082
sources which is almost a factor two larger than the sample of 599 sources used by ~\citet[][]{Farnes2014ApJ...795...63F}. Among them 476 have $\alpha \leq -0.7$ and 315 with $\alpha
\geq -0.3$. The CPDFs of $\vert RRM \vert$ for both these samples, after
sub-classifying them in n(Mg~{\sc ii}) = 0 and n(Mg~{\sc ii}) $>$ 0
sub-samples are shown in Fig.~\ref{aplha_cuml_rrm1}. As can be seen from this figure, for the subclass of flat spectrum sources, the $\vert RRM \vert$ consistently appears to be higher for n(Mg~{\sc ii}) $>$ 0 in comparison to the case of n(Mg~{\sc ii}) = 0, though the $P_{null}$ of $24\%$ based on AD-test is still high. However for the subsample of steep spectrum sources, the value of $P_{null} = 69\%$ is found to be much higher though the difference is not significant.

The values of $\sigma_{rrm}^{md}$, as listed in the lower part of Table~\ref{table:sigma}, is also found to be consistent with the above conclusion. The $\sigma_{md}^{ex}$ of 9.8$\pm2.8$ \rrm~ for flat spectrum sources suggests the correlation at $\sim 3.5 \sigma$ among the presence of \mgii absorbers and $\sigma_{rrm} $. This correlation is almost absent in the subclass of steep spectrum sources (with excess at $\sim 0.5 \sigma$ level). This result is found to be consistent with the similar $3.5\sigma$ correlation reported by ~\citet[][]{Farnes2014ApJS..212...15F}.\par
We also note that the result based on SD method for flat 
spectrum sources is also consistent to that derived using the MADFM, though these two method depart for the case of subclass of steep spectrum sources (e.g., see Table~\ref{table:sigma}), perhaps due to few outlier point in this subclass. Given the fact that MADFM is more resistant to the outlier points compared to the SD, the result based on MADFM can be considered to be more robust.
%
% %\hline
%

\subsection{Correlation of RRM with polarization percentage}

\label{co_rm_p}

\begin{figure}
  \epsfig{figure=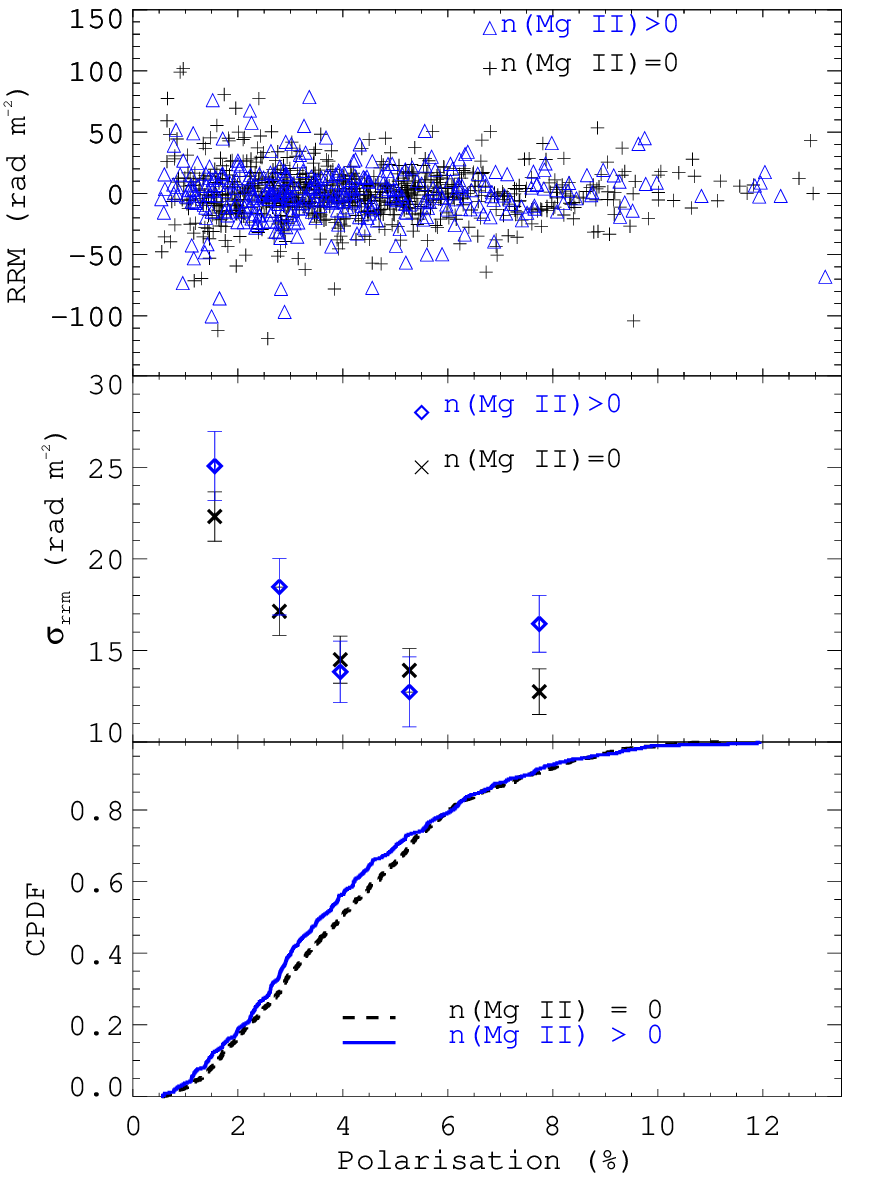,height=15.cm,width=9.cm,angle=0}
\caption{\emph{Upper panel:} Distribution of RRM with the polarization percentage, $p$, for sources with (blue triangle) and without (black plus) \mgii absorbers along the quasar's sightline. \emph{Middle panel:} The plot of broadening of RRM (i.e., $\sigma_{rrm}^{md}$) with the polarization percentage for subsamples with (blue diamond) and without (black cross) \mgii absorbers is shown. Their values of Pearson correlation coefficient, ($\rho_{p})$ is $-0.62$ and $-0.87$, respectively. \emph{Lower panel:} The CPDFs of $p$, for subsample with (blue solid line) and without (black dashed line) \mgii  absorbers. }
\label{fig:polarization}
\end{figure}

\begin{figure}
  \epsfig{figure=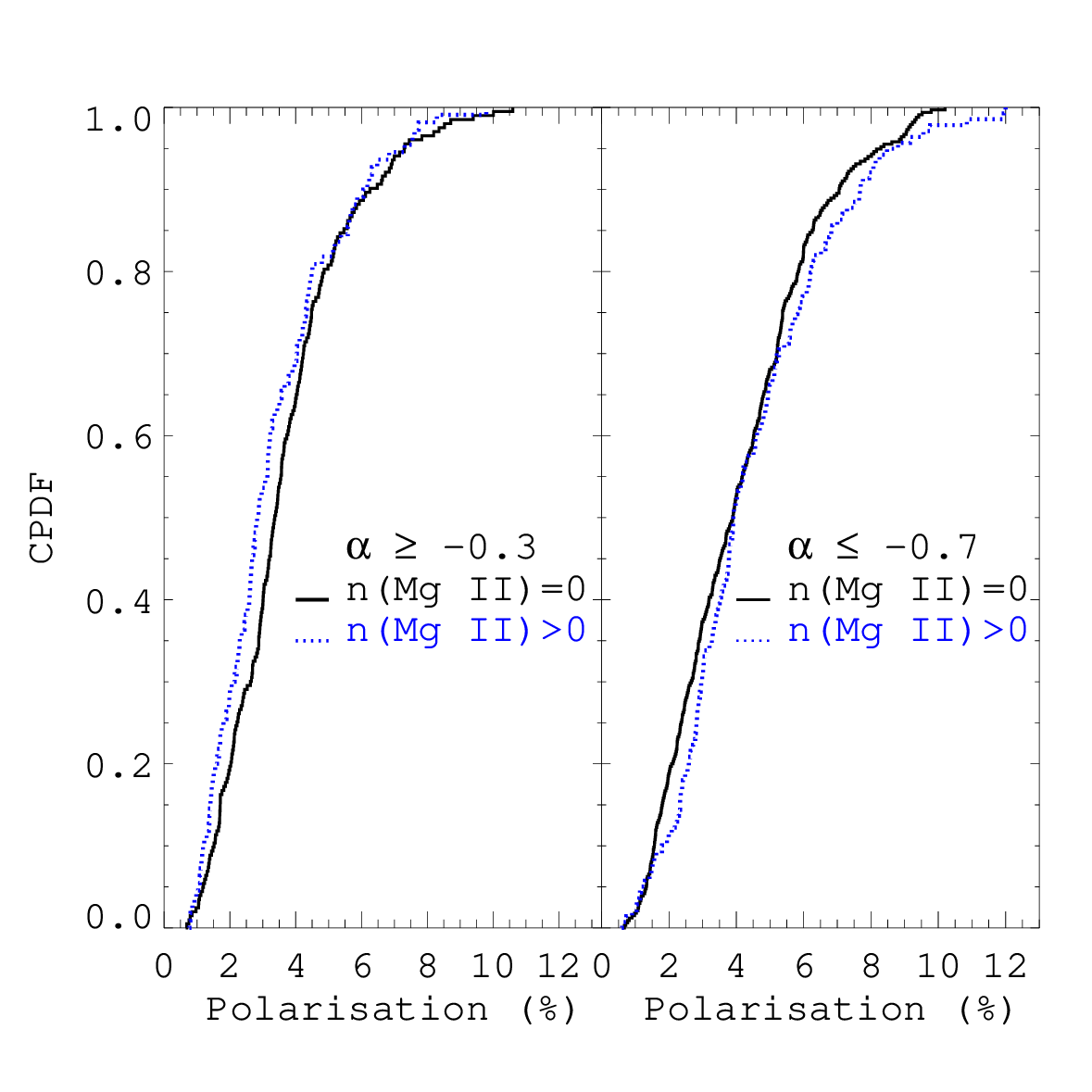,height=8cm,width=9cm,angle=0}
\caption{\emph{Left panel:} The CPDFs of the polarization percentage, $p$, for the quasar with no absorber (black line), with absorbers (blue dotted line) for $\alpha \geq -0.3$. \emph{Right panel:} same as left, but for steep spectrum quasars sample with $\alpha \le -0.7$.}
\label{aplha_cuml_rrm}
\end{figure}

The polarisation percentage, ($p$) measurement were available for all
our 1132 sources. To test the hypothesis that intervening absorber
might also lead to depolarisation based on mechanisms such as differential
Farady rotation and/or due to the beam depolarisation ~\citep[e.g., see][]{burn1966,sokoloff1998,flecther2004,Kim2016ApJ...829..133K,2018arXiv181003638K}, we first plotted the RRM versus $p$ in the upper
panel of Fig.~\ref{fig:polarization}, which show a trend of decrease in RRM scatter with the increase in $p$ values. To quantify this trend we have also plotted the $\sigma_{rrm}^{md}$ (based on MADFM)
in various bins of $p$ as shown in the middle panel of Fig.~\ref{fig:polarization}. From this figure an anti-correlation between $\sigma_{rrm}^{md}$ and $p$ can be noticed in the subsample
of sightlines having \mgii absorbers as well as in the subsample of
sightlines without any such absorbers, with a Pearson correlation
coefficient, $\rho_{p}$, of $-0.62$ and $-0.88$, with significance level of 75\% and 95\%, respectively. In the bottom panel of Fig.~\ref{fig:polarization}, we have also plotted the CPDF of polarization percentage. As can be seen from this figure, the fractional polarization for subsample with n(Mg~{\sc ii})=0 seem to be nominally higher in a systematic manner as compared to the  subsample with n(Mg~{\sc ii})$>$0. This nominal difference due to the absorbers is also revealed by the AD-test which shows that the $P_{null}$ among these two distribution is 12\%.

Additionally, we also repeated our above analysis separately in the flat and steep spectrum sources, by comparing their respective CPDFs of percentage polarisation  based on subsamples with and without \mgii absorbers as shown in Fig.~\ref{aplha_cuml_rrm}. It can be noticed from this figure that the difference (among subsample with and without \mgii) might be more in the subclass of flat spectrum as compared to the subclass of the steep spectrum sources, though in neither subclass it is clearly evident. This is also corroborated by the AD-test which shows that the  P$_{null}$ is only 5\% and 16\% for the former and latter subclass, respectively.

The differences between subsamples with and without \mgii absorbers not being significant (both in full as well as in subsample splitting based on spectral indices), suggest that dominant contribution towards depolarization is due to the local environment of the background quasar instead of the intervening galaxy~\citep[see also][]{Farnes2014ApJ...795...63F}.

\begin{figure}
\epsfig{figure=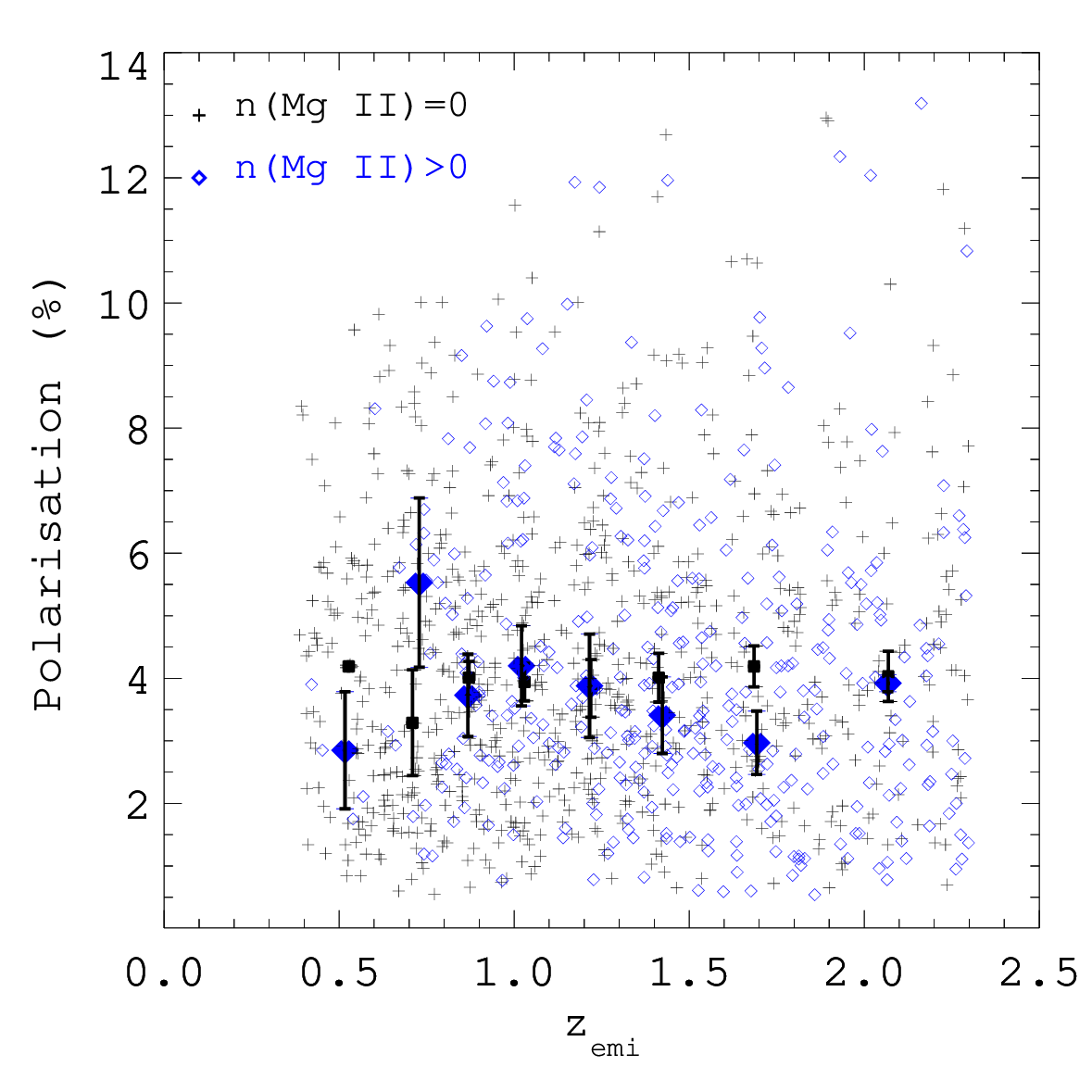,height=9cm,width=9cm,angle=0}
\caption{ The distribution of polarization percentage ($p$) with emission redshift of the quasar, $z_{emi}$, for subsample without \mgii absorber (black plus), and for subsample with \mgii absorbers (blue diamond). The thick point with error are the corresponding median value within the redshift bins size of 0.238. }

 \label{aplha_cuml_redshift}
\end{figure}

\begin{figure}

   \epsfig{figure=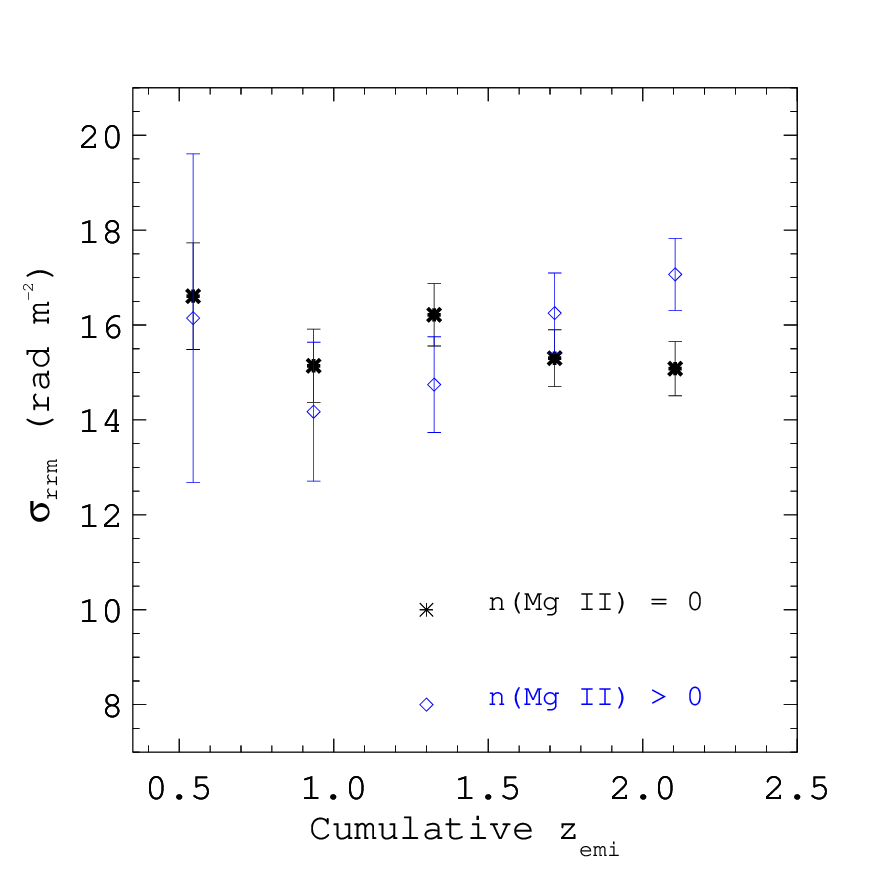,height=9.0cm,width=9.0cm,angle=0}
\caption{The plot of brodening of RRM (i.e., $\sigma^{md}_{rrm}$) both for subsample with (blue diamond) and without (black star) \mgii systems in a common cumulative bins of $z_{emi}$.}
\label{fig:rrm_absorber_redshift}
\end{figure}

%%%%%%%%%%%%%%%%%%%%%
\subsection{The redshift evolution of RRM and fractional polarisation}
\label{red_evol}
Any trend of the broadening of RRM distributions and value of polarisation with the redshift could be a useful tracer of the evolution of galactic magnetic fields. For this we first plotted in 
Fig.~\ref{aplha_cuml_redshift}  the  value of the percentage  polarisation ($p$) verses the $z_{emi}$. As can be seen from this plot, $z_{emi}$ and $p$ seem to be uncorrelated both in the case of subsamples with \mgii absorbers as well as the ones without. This is also evident on the basis of the over-plot of the median value of $p$ within a bin of $z_{emi}$ (using bin size of $\sim 0.238$), which appears almost similar for subsamples with and without \mgii absorbers within the error-bar. Here the error-bar within the bin has been computed by the quadratic sum of two sources of errors, viz., the resultant of (i) individual error on $p$ values within a bin and (ii) the statistical error computed as an r.m.s deviation  within the bin around the median value  of $p$ (i.e., $\sqrt{variance/(N-1)}$). From this figure, it is clear that as far as the value of percentage  polarisation and its redshift evolution  is concerned the local environment dominates in comparison to the impact of intervening galaxy (if any).\par

Further, it is also tempting to look for any trend of $\sigma_{rrm}$ and its excess at different redshift bins, as it could be a useful tracer of the evolution of galactic magnetic fields in the high redshift galaxies. However, we noticed that with such binning in redshift space the estimate of broadening of RRM distributions could easily dominate due to the low number statistics, especially while computing the excess among subsample with and without \mgii absorbers. As an alternative, we show in Fig~\ref{fig:rrm_absorber_redshift}, the plot of $\sigma_{md}^{rrm}$ over the cumulative bin of $z_{emi}$, where in the successive bin data from the higher redshift is consecutively being included to see any evolution in the $\sigma_{md}^{rrm}$,  beside excluding the possibility of low-number statistics. 
As can be seen from this figure that over the redshift range of 0.38 to 2.3 of our sample the evolution in $\sigma_{md}^{rrm}$ (or diffrence among samples in a bin with and without \mgii system) with redshift does not seem to be very strong withing the error-bars. Perhaps much larger data-set may be able to distinguished any tentative diffrence among the two subsample at various redshift bins.

\section{Discussion and Conclusions}

As pointed out in Sect.~\ref{introduction}, many past studies
have analysed the impact of intervening systems on the RM
distributions. Here we have carried out a similar study based on a sample size (containg 1132 sources based on SDSS 7, 9, 12 \& 14) which is about twice as large as compared to previous large samples of 539  and 599 sources by 
~\citet[][]{Joshi2013MNRAS.434.3566J} and ~\citet[][]{Farnes2014ApJ...795...63F} (e.g., see Section~\ref{introduction}) respectively.

Special care has been taken to separate the sightlines, with and without \mgii absorbers by supplementing our automated search with a visual check of each system (see Sect.~\ref{identification}). We also employed the new method based on MADFM to compute the broadening (i.e., $\sigma_{rrm}^{md}$) in RRM distributions, which is found to be more
resistant to the outliers as compared to the traditonally applied SD
method. The excess in $\sigma_{rrm}$ based on these two methods is 
%$\sigma^_{rrm}^{md}$ and $\sigma^{ex}_{rrm}^{SD}$ is
found to be 8.0$\pm 1.9$~\rrm~  and 7.5$\pm 2.0$~\rrm~, respectively, using the 352 sightlines with n(Mg~{\sc ii})$>$0 and the 780 sightlines with n(Mg~{\sc ii}) =0. This shows that the galaxies responsible for these intervening \mgii systems do have a contribution in the RRM, which is evident at the confidence level of 4.2$\sigma$ and 3.7$\sigma$ based on MADFM and SD method, respectively. \par

Further, comparing our above mentioned 3.7$\sigma$ excess, using $\sigma^{ex}_{rrm}$= \textbf{$7.5\pm 2.0$},  based on the SD method with that of the estimate of the excess of 1.7$\sigma$ by ~\citet[][]{Joshi2013MNRAS.434.3566J} using the same SD method with  $\sigma^{ex}_{rrm}$= $8.11\pm4.83$, implies that these  are consistent with each other. This is apart from a gain in precision (about factor 2) in our measurement mainly due to the  better statistics based on our enlarged sample size (being about factor two larger).
Similarly, an excess is also reported by many past studies  ~\citep[e.g.,][]{Kronberg2008ApJ...676...70K,Bernet2008Natur.454..302B,Bernet2010ApJ...711..380B,Farnes2014ApJ...795...63F,mao2017}).

For instance, ~\citet[][]{Farnes2014ApJ...795...63F} reported  an excess, based on a method of computing difference in the median value of RM distributions which is found to be 6.9$\pm$1.7~\rrm~ (i.e., at 4$\sigma$), consistent with the trend found in our analysis. 
The significant excess in RRM reported by  ~\citet[][]{Bernet2008Natur.454..302B} ~\citep[see also,][]{Bernet2010ApJ...711..380B}, has been arrived at by using RRM measurements at 6cm wavelength (using 76 sources) unlike all the above discussed studied based on RRM at 21cm wavelength. As proposed by ~\citet[][]{Bernet2012ApJ...761..144B}, the $\sigma^{ex}_{rrm}$ can be wavelength dependent with more dilution at 21cm based on their inhomogeneous Faraday screen models. In view of the fact that we find an excess in our analysis suggests that the RRM contribution of intervening galaxies is also found to be significant at 21cm wavelength as at 6cm wavelength, though it can also be partly affected by a nominal contribution of the above propose inhomogeneous Faraday screen phenomena.

Additionally, similar to the ~\citet[][]{Farnes2014ApJ...795...63F}, we also searched for any bias due to the radio emission morphology such as compact core dominated and extended lobe dominated sources by making a split based on their spectral index, $\alpha$, (where $F_\nu \propto \nu^{\alpha}$) namely, $\alpha \ge -0.3$ and $\alpha \le -0.7$, respectively. 
 
As detailed in Sect.~\ref{result2}, we found that excess of $\sigma_{rrm}$ in flat sources does indeed correlate at high significance of 3.5$\sigma$ level with the presence  of \mgii absorber, but such correlation is almost absent in steep spectrum sources (see lower part of Table~\ref{table:sigma}). This is found to be consistent with that of ~\citet[][]{Farnes2014ApJ...795...63F} where they have reported the $\sim 3.5\sigma$ difference. However, we also noted here that their result was based on $P_{null}$ estimated using KS-test by comparing the empirical cumulative-distribution functions  of their two RM distributions (with and without \mgii absorbers) unlike our case  where we have computed the $\sigma^{ex}_{rrm}$. This  difference among the methods used, may also be a possible reason that in spite of the larger sample in our case the expected relative improvement in precision of the above correlation is almost similar to the relatively smaller sample (almost by factor of 2) used in the  ~\citet[][]{Farnes2014ApJ...795...63F} studies.

Other important additional advantages of our enlarged sample are that
we could divide the sample in various bins of redshift and fractional
polarisation, to test any evolution of $\sigma_{rrm}$ with these
parameters. Our analysis indicate an anti-correlation among $\sigma^{md}_{rrm}$ and the  fractional polarisation
with Pearson correlation coefficient of $-0.62$ and
$-0.87$ for the subsample with and without \mgii absorbers (see Sect.~\ref{co_rm_p}, and Fig.~\ref{fig:polarization}). The higher $\sigma^{md}_{rrm}$ found at low fractional polarisation bin, might represent more randomisation and hence more decrements in the fractional polarisation, leading to the above observed decrements of $\sigma^{md}_{rrm}$ by about 10 \rrm\ over a range of fractional polarisation from 0 to 8\%  (see the middle panel of Fig.~\ref{fig:polarization}).
The similarity of the anticorrelation separately in subsamples with and without \mgii absorbers shows that the intervening absorbers have negligible impact on the fractional polarization (at least in 21cm wavelength). 
The impact of morphology also seems to be negligible, based on our comparison of the CPDF of subsample with and without \mgii absorbers separately for the subclass of flat and steep-spectrum sources (e.g., see Fig.~\ref{aplha_cuml_rrm}). The possibility of any dominant role 
due to the IGM based on mixing of the sources with diffrent emission redshifts in a given polarisation bin, is also unlikely as we do not see any evolution of the fractional
polarisation with $z_{emi}$ both for subsample with as well as without \mgii systems (see Fig.~\ref{aplha_cuml_redshift}). Thus our analysis suggests that the dominant factor to determine the level of  polarization is based on local environment instead the intervening galaxies. This is also found consistent with the studies of ~\citet[][]{Farnes2014ApJ...795...63F}.
However the studies by ~\citet[][]{Kim2016ApJ...829..133K} suggest  that the intervening systems are strongly associated with depolarization, 
though they use a different technique of rotation measure synthesis over a rather smaller sample size (49 unresolved quasars).

For the RRM excess $\sigma^{ex}_{rrm}=8.0\pm1.9$~\rrm~(using MADFM) obtained for our subsample with (352 sources) and without (780 sources) \mgii systems 
(see Table~\ref{table:sigma}, top row), lead to typical estimate of  $\langle B_\parallel\rangle$ $\sim 1.3\pm 0.3~\mu G$ for the intervening galaxies responsible for these \mgii absorbers with the median redshift of 0.92 (e.g., see Sect.~\ref{result1}). This is found to be consistent with the recent similar estimate by ~\citet[][]{Farnes2014ApJ...795...63F} of magnetic field of about $1.8\pm 0.4~\mu G$ based on the exesss of $6.9\pm1.7$ ~\rrm~ in the RM value (using 599 sources) calculated based on the difference among the median value of RM distributions. The typical value of magnetic field estimated by~\citet[][]{Bernet2008Natur.454..302B} is about $10\mu G$, using the RRM excess based on the RRM measurements at 6cm (using 76 sources). Our result also supports these previous findings that high-z galaxy contain large enough magnetic field to be observable as an significant effect on the measured RRM value.
We would like to point out that the strength of the average magnetic fields reported here is derived based on the integrated values traced by \mgii absorbers having range of  impact parameter from the center of galaxies. Hence this should be treated as an average value of the magnetic field over the galactic scale.

On the other-hand, the typical estimate of magnetic field based on the energy equipartition (pressure equilibrium) in the galactic systems also varies from 1-10$\mu$G~\citep[e.g., see][and reference therein]{2005AN....326..414B}, by assuming the pressure balance between the cosmic rays with that of the galactic magnetic field. Similar estimates based on the radio-far infrared correlation varied over a range of 8 to 23$\mu G$ in redshift upto 0.05~\citep[e.g., see][]{2018MNRAS.477.3552O} and references therein~\citep[][]{dominik2016,basu2017,mao2017}, which are significantly higher than our estimate above based on RRM excess (i.e., $\sigma^{ex}_{rrm}$). However, we also note here that these studies at very low redshift consist of relatively small sample of peculiar galaxies such as one with high star formation etc, can perhaps have high magnetic field. This could also arise at low-z based on amplification of galactic magnetic filed due to small-scale turbulent dynamo action ~\citep[][]{Bhat2013MNRAS.429.2469B,2018MNRAS.477.3552O}. Nonetheless such observational constrain as obtain above together with comparative studies of RRM for sightline with and without \mgii systems will prove to be important for building any model of galactic magnetism at high-redshift, especially in the context of our understanding of the galaxy formation.

  Further improvements on these aspects will be possible with an enlarged sample having both optical spectra as well as radio polarisation measurements. Surveys, such as the square kilometer array (SKA) as well as future optical data release like of SDSS can play crucial roles in these studies.

\section*{Acknowledgments}
We thank the anonymous referee for the constructive comments on our manuscript. 
SM, HC and TRS acknowledge the facilities at IRC, University of Delhi. SM and TRS also thank ARIES for the hospitality. TRS and HC acknowledge the project grant from SERB, Govt. of India (EMR/2016/002286). The research of SM is supported by UGC, Govt. of India under the UGC-JRF scheme (Sr.No. 2061651305  Ref.No: 19/06/2016(I) EU-V).

 \par
 Funding for the SDSS and SDSS-II has been provided by the Alfred P.
 Sloan Foundation, the Participating Institutions, the National
 Science Foundation, the U.S. Department of Energy, the National
 Aeronautics and Space Administration, the Japanese Monbukagakusho,
 the Max Planck Society, and the Higher Education Funding Council for
 England. The SDSS Web Site is http://www.sdss.org/. The SDSS is
 managed by the Astrophysical Research Consortium for the
 Participating Institutions. The Participating Institutions are the
 American Museum of Natural History, Astrophysical Institute Potsdam,
 University of Basel, University of Cambridge, Case Western Reserve
 University, University of Chicago, Drexel University, Fermilab, the
 Institute for Advanced Study, the Japan Participation Group, Johns
 Hopkins University, the Joint Institute for Nuclear Astrophysics, the
 Kavli Institute for Particle Astrophysics and Cosmology, the Korean
 Scientist Group, the Chinese Academy of Sciences (LAMOST), Los Alamos
 National Laboratory, the Max-Planck-Institute for Astronomy (MPIA),
 the Max-Planck-Institute for Astrophysics (MPA), New Mexico State
 University, Ohio State University, University of Pittsburgh,
 University of Portsmouth, Princeton University, the United States
 Naval Observatory, and the University of Washington.

%\lastpage
\bibliography{references}
\label{lastpage}
\end{document}